\documentclass{article}
\usepackage{spconf,amsmath,graphicx}
\usepackage{booktabs}
\usepackage{hyperref}
\usepackage{tikz}
\usepackage{pgfplots}
\usepackage{soul}
\usepackage{multirow}
\usepackage{caption}
\usepackage{subcaption}

\title{End-to-End Speech Recognition from Federated Acoustic Models}
\name{Yan Gao$^1$, Titouan Parcollet$^{2,1}$, Salah Zaiem$^3$, Javier Fernandez-Marques$^4$, Pedro P. B. de Gusmao$^1$, Daniel J. Beutel$^{1,5}$, and Nicholas D. Lane$^{1}$}
\address{$^1$University of Cambridge, $^2$Avignon University\\ $^3$Telecom Paris, $^4$University of Oxford, $^5$Adap GmbH}

\begin{document}
%
\maketitle
\begin{abstract}
Training Automatic Speech Recognition (ASR) models under federated learning (FL) settings has attracted a lot of attention recently. However, the FL scenarios often presented in the literature are artificial and fail to capture the complexity of real FL systems. In this paper, we construct a challenging and realistic ASR federated experimental setup consisting of clients with heterogeneous data distributions using the French and Italian sets of the CommonVoice dataset, a large heterogeneous dataset containing thousands of different speakers, acoustic environments and noises. We present the first empirical study on attention-based sequence-to-sequence End-to-End (E2E) ASR model with three aggregation weighting strategies -- standard FedAvg, loss-based aggregation and a novel word error rate (WER)-based aggregation, compared in two realistic FL scenarios: \textit{cross-silo} with $10$ clients and \textit{cross-device} with $2$K and $4$K clients. Our analysis on E2E ASR from heterogeneous and realistic federated acoustic models provides the foundations for future research and development of realistic FL-based ASR applications.
\end{abstract}
\begin{keywords}
End-to-end ASR, federated learning
\end{keywords}
\section{Introduction}

Deep neural networks are now widely adopted in state-of-the-art (SOTA) ASR systems \cite{kumar2018survey}. This success mostly relies on the centralised training paradigm where data needs first to be gathered in one single dataset before it can be used for training \cite{hannun2014deep, amodei2016deep, soltau2016neural}. Such approach has a few clear benefits including fast training and, the ability to sample data in any preferred way due to the complete data visibility. However, recent concerns around data privacy along with the proliferation of both powerful mobile devices and low latency communication technologies (e.g.\ 5G), distributed training paradigms such as FL begin to receive more attention.  

In FL, training happens at the source and training data is never sent to a centralised server. In a typical FL scenario each participating client/device receives a copy of an initial model and separately trains its model on its local data. This process generates a set of weight updates that are then sent to a server, where updates are aggregated. This process is repeated for several rounds\cite{mcmahan2017communication, konevcny2016federated, kairouz2019advances}. 
Being able to harvest information from thousands of mobile devices without having to collect users' data makes federated and on-device training of ASR systems a feasible and an attractive alternative to traditional centralised training~\cite{konevcny2016federated}, whilst offering new opportunities to advance ASR quality and robustness given the unprecedented amount of user data directly available on-device.  
For example, such data could be leveraged to better adapt the ASR model to the users' usage, or improve the robustness of models to realistic and low resources scenarios \cite{kairouz2019advances}.

Despite the growing number of studies applying FL on speech-related tasks \cite{hard2020training, leroy2019federated, dimitriadis2020federated, cui2021federated, granqvist2020improving}, very few of these have investigated its use for E2E ASR. Properly training E2E ASR models in a realistic FL setting comes with numerous challenges. First, it is notoriously complicated to train a deep learning model with FL on non independent and identically distributed data (non-IID)~\cite{kairouz2019advances, zhao2018federated, sattler2019robust} and on-device speech data is extremely non-IID by nature (e.g. different acoustic environments, words being spoken, languages, microphones, amount of available speech, etc.). Second, state-of-the-art E2E ASR models are computationally intensive and potentially not suited for on-device training phases of FL. Indeed, the latest ASR systems rely on large Transformers \cite{mohamed2019transformers, zeyer2019comparison}, Transducers \cite{mohri2002weighted, battenberg2017exploring} or attention sequence-to-sequence (Seq2Seq) models \cite{kim2017joint, chiu2018state} that process high-dimensional acoustic features. 
Finally, E2E ASR training is difficult and very sensitive during early stages of optimisation due to the complexity of learning a proper alignment between the latent speech representation and the transcription. These three traits make it very difficult to train ASR models completely from scratch \cite{rosenberg2017end,bansal2019pre}.

To the best of our knowledge, existing works on FL for ASR typically approach these challenges by relinquishing few constraints of the environmental protocol. This in turn results in their experimental settings being still far away from the conditions in which a FL ASR would need to function. In fact, many works\cite{guliani2021training, dimitriadis2020federated} are evaluated on unrealistic datasets (\textit{w.r.t} the FL scenario) such as LibriSpeech (LS) \cite{7178964}, which only contains recordings from selected speakers reading books in a controlled setting without background noise. Authors in~\cite{cui2021federated} introduce a client-based adaptive training of a HMM-DNN based ASR system for \textit{cross-silo} FL, a specific setup which considers only a reduced number of clients with large amounts of homogeneous data, thus simplifying the complexity of real FL setups where data is non-IID.
Then, \cite{dimitriadis2020federated} propose a federated transfer learning platform with improved performance using enhanced federated averaging and hierarchical optimization for E2E ASR, alleviating the aforementioned alignment issue but only evaluating it on LS.

In this work, we motivate the need to move away from clean speech corpora for evaluating FL-based ASR systems. We investigate FL models in a more realistic setting with the French Common Voice (CV) dataset \cite{ardila2019common}, which provides a large, heterogeneous and uncontrolled set of speakers who used their own devices to record a given set of sentences; naturally fitting to FL with various users, acoustic conditions, microphones and accents. We evaluate both a \textit{cross-silo} and a \textit{cross-device} (i.e. large number of clients with few naturally non-IID data) FL setups while training a SOTA E2E ASR system. In particular, we compare three different weighting strategies during models aggregation. Our contributions are: 
\begin{enumerate}
    \item Quantitatively compare LibriSpeech to Common Voice towards a realistic FL setup to highlight the need for a shift in the evaluation of FL-based ASR models. 
    \item Present the first study on attention-based Seq2Seq E2E ASR model for FL scenarios. Our setup investigates challenges previously overlooked by others, such as extremely heterogeneous recording conditions.
    \item Evaluate both \textit{cross-silo} and \textit{cross-device} FL with up to 4K clients on the naturally-partitioned and heterogeneous French and Italian subsets of Common Voice.
    \item A first adapted aggregation strategy based on WER to integrate the specificity of ASR to FL. 
    \item Release the source code using Flower~\cite{beutel2020flower} and SpeechBrain~\cite{SB2021} to facilitate replication and future research\footnote{\url{github.com/yan-gao-GY/Flower-SpeechBrain}}. 
\end{enumerate}

\section{End-to-end Speech Recognizer}
The considered E2E ASR system relies on the wide spread joint connectionist temporal classification (CTC) with attention paradigm~\cite{kim2017joint}. This method combines a Seq2Seq attention-based model~\cite{bahdanau2016end} with the CTC loss~\cite{graves2014towards}. 

A typical ASR Seq2Seq model includes three modules: an encoder, a decoder and an attention module. Given a speech input sequence (i.e. speech signal or acoustic features) $\textbf{x} = [x_1, ..., x_{T_x}]$ with a length $T_x$, the \textit{encoder} first converts it into a hidden latent representation $\textbf{h}^e = [h^e_1 , ..., h^e_{T_x}]$. Then the \textit{decoder} attends to the encoded representation $\textbf{h}^e$ combined with an attention context vector $c_t$ from the attention module. This produces the different decoder hidden states $\textbf{h}^d = [h^d_1 , ..., h^d_{T_y}]$, with $T_y$ corresponding to the length of the target sequence $\textbf{y}$. In a speech recognition scenario $T_x > T_y$.

The standard training procedure of the joint CTC-Attention ASR pipeline is based on two different losses over a dataset $S$. First, the CTC loss is derived with respect to the prediction obtained from the encoder module of the Seq2Seq model:

\begin{equation}
\mathcal{L}_{CTC} = - \sum_S \log p(\textbf{y}|\textbf{h}^e),
\label{eq1}
\end{equation}
Second, the attention-based decoder is optimised following the cross entropy (CE) loss:
\begin{equation}
\mathcal{L}_{CE} = - \sum_S \log p(\textbf{y}|\textbf{h}^d).
\label{eq2}
\end{equation}
The losses are combined with a hyperparameter $\mu\in[0,1]$ as:

\begin{equation}
\mathcal{L} = \mu \mathcal{L}_{CE} + (1-\mu)\mathcal{L}_{CTC}.
\label{eq3}
\end{equation}

In practice the CTC loss facilitates the early convergence of the system due its monotonic behavior while the attentional decoder needs to first figure out where to attend in the hidden representation of the entire input sequence. 

\section{Federated Training of Acoustic Models}

The process of training an E2E acoustic model using federated learning follows four steps: 1) Following~\cite{dimitriadis2020federated}, model weights are initialised with a pre-training phase on a centralised dataset; 2) The centralised server samples $K$ clients from a pool of $M$ clients and uploads to them the weights of the model. 3) The clients train the model for $t_{local}$ local epochs in parallel based on their local user data and send back the new weights to the server. 4) The server aggregates the weights and restart at step 2. This procedure is executed for $T$ rounds until the model converges on a dedicated validation set (e.g. local to each client or centralised). 

\subsection{Federated Optimisation}

For each training round, each client $k \in K$, containing  $n_k$ audio samples, runs $t \in [0, t_{local}]$ iterations with learning rate $\eta_{l}$ to locally update the model based on Eq. \ref{eq3},

\begin{equation}
w_{t+1}^{(k)} = w_t^{(k)} - \eta_{l}\tilde{g_k},
\label{eq4}
\end{equation}

with $w_k$ the local model weights iof client $k$, and $\tilde{g_k}$ the average gradient over local samples. After training for $t_{local}$ local epochs in the global round $T$, the updated weights $w^{(k)}_T$ of the client $k$ are sent back to the server. Then, the local gradient $g^{(k)}_T$ is computed as:

\begin{equation}
g^{(k)}_T = w^{(k)}_T - w_{T - 1}.
\label{eq5}
\end{equation}
Then, the gradients from all clients are aggregated as follows:

\begin{equation}
 \Delta_T = \sum_{k = 1}^K\alpha^{(k)}_{T}g^{(k)}_T, 
\label{eq6}
\end{equation}

where $\alpha^{(k)}_T$ denotes different weighting strategies described in Section \ref{sec:weight}. The updated global model weights $w_T$ are computed with a server learning rate $\eta_s$ according to:

\begin{equation}
 w_T = w_{T - 1} - \eta_s\Delta_T,  
\label{eq7}
\end{equation}

During FL training, especially with heterogeneous data, the global model may deviates away from the original task or simply not converges~\cite{kairouz2019advances, zhao2018federated, sattler2019robust}, and therefore lead to performance degradation. To alleviate this issue, and motivated by~\cite{dimitriadis2020federated}, we propose an additional training iteration over a small batch of held-out data on the server, after the standard model update procedure with Eq.\ref{eq7}. 

\subsection{Weighting Strategies}
\label{sec:weight}

Federated Averaging (FedAvg)~\cite{mcmahan2017communication} is a popular~\cite{li2020federated,horvath2021fjord,qiu2021look,hard2020training, leroy2019federated, guliani2021training} aggregation strategy by which model updates from each client are weighted by $\alpha^{(k)}_T$, the ratio of data samples in each client over the total samples utilized in the round:

\begin{equation}
\alpha^{(k)}_T = \frac{n_k}{\sum_{k=1}^K n_k}, 
\label{eq8}
\end{equation}

In realistic FL settings with heterogeneous client data distribution, some clients may contain data that is skewed and not representative of the global data distribution (e.g. audio samples with different languages or multiple speakers). As a result, the aggregated model might simply not converge if such clients have proportionally more training samples than others. For instance, in our experiments, all attempts to train an ASR system from scratch failed due to this issue requiring a prior pre-training phase of the acoustic model. 
Second, clients containing low quality data would introduce unexpected noise into the training process (e.g. extreme noise in the background). Either scenario could lead to model deviation in the aggregation step, which can not be solved via the standard FedAvg weighting method (Eq. \ref{eq8}). A potential solution, instead, is to use the averaged training loss as a weighting coefficient, thus reflecting the quality of the locally trained model. Intuitively, higher loss would indicate that the global model struggles to learn from the client's local data. More precisely, we compute the weighting with the \textit{Softmax} distribution obtained from the training loss from Eq. \ref{eq3}. Eq. \ref{eq8} is modified as follows:

\begin{equation}
\alpha^{(k)}_T = \frac{\exp{(- \mathcal{L}_k)}}{\sum_{k=1}^K \exp{(- \mathcal{L}_k)}}.  
\label{eq9}
\end{equation}

In the context of ASR, WER is commonly used as the final evaluation metric for the model instead of the training loss. We therefore propose a WER-based weighting strategy for aggregation. This approach utilizes the values $(1 - wer)$ obtained on the validation set as weighting coefficients $\alpha^{(k)}_T$:

\begin{equation}
\alpha^{(k)}_T = \frac{\exp{(1 - wer_k)}}{\sum_{k=1}^K \exp{(1 - wer_k)}}.  
\label{eq10}
\end{equation}

In this way, we directly optimise the model towards the relevant metric for speech recognition.

\section{Common Voice as a realistic FL setup}

In this section we first present the Common Voice (CV) dataset used for the FL experiments. Then, we quantitatively demonstrate that CV is a much more adapted corpus to advance FL research than LibriSpeech (LS), motivating the need for a shift in the standard evaluation process. 

\subsection{Common Voice dataset}
\label{sec:dataset}
Both the French and Italian subsets of CV dataset (version 6.1)~\cite{ardila2019common} are considered. Utterances are obtained from volunteers recording sentences all around the world, and in different languages, from smartphones, computers, tablets, etc. The French set contains a total of $328$K utterances ($475$ hours in total) with diverse accents which were recorded by more than $10$K French-speaking participants. The train set consists of $4212$ speakers ($425.5$ hours of speech), while both validation and test sets contain around $24$ hours of speech from $2415$ and $4247$ speakers respectively. The Italian set, on the other hand, is relatively small, containing $89$, $21$ and $22$ hours of Italian training ($748$ speakers), validation ($1219$ speakers) and test ($3404$ speakers) data.

\subsection{Setup analysis and LibriSpeech comparison} 
We argue that CV is closer to natural federated learning conditions than LS as much stronger variations are observed both intra- and inter-clients. While CV is a crowd-sourced dataset containing thousands of different acoustic conditions, microphones and noises, LS is a heavily controlled studio-quality corpus. The latter has been used by most research on FL ASR.
We compare both datasets at three levels:

\textbf{Low-level signal features.} The selected features should be more descriptive of the background and recording conditions than speaker identity, as this is investigated when analysing clustering purity. Hence, we will consider: Loudness as it is highly linked to the microphone and the recording distance; the log of the Harmonicity to Noise Ratio (logHNR) as a proxy indicator of background noise; Permutation Entropy (PE) as it has been successfully used for microphone identification purposes\cite{baldini2020}.

\begin{figure}[!t]
\centering
\begin{subfigure}{.45\textwidth}
  \centering
   \includegraphics[width=.7\linewidth]{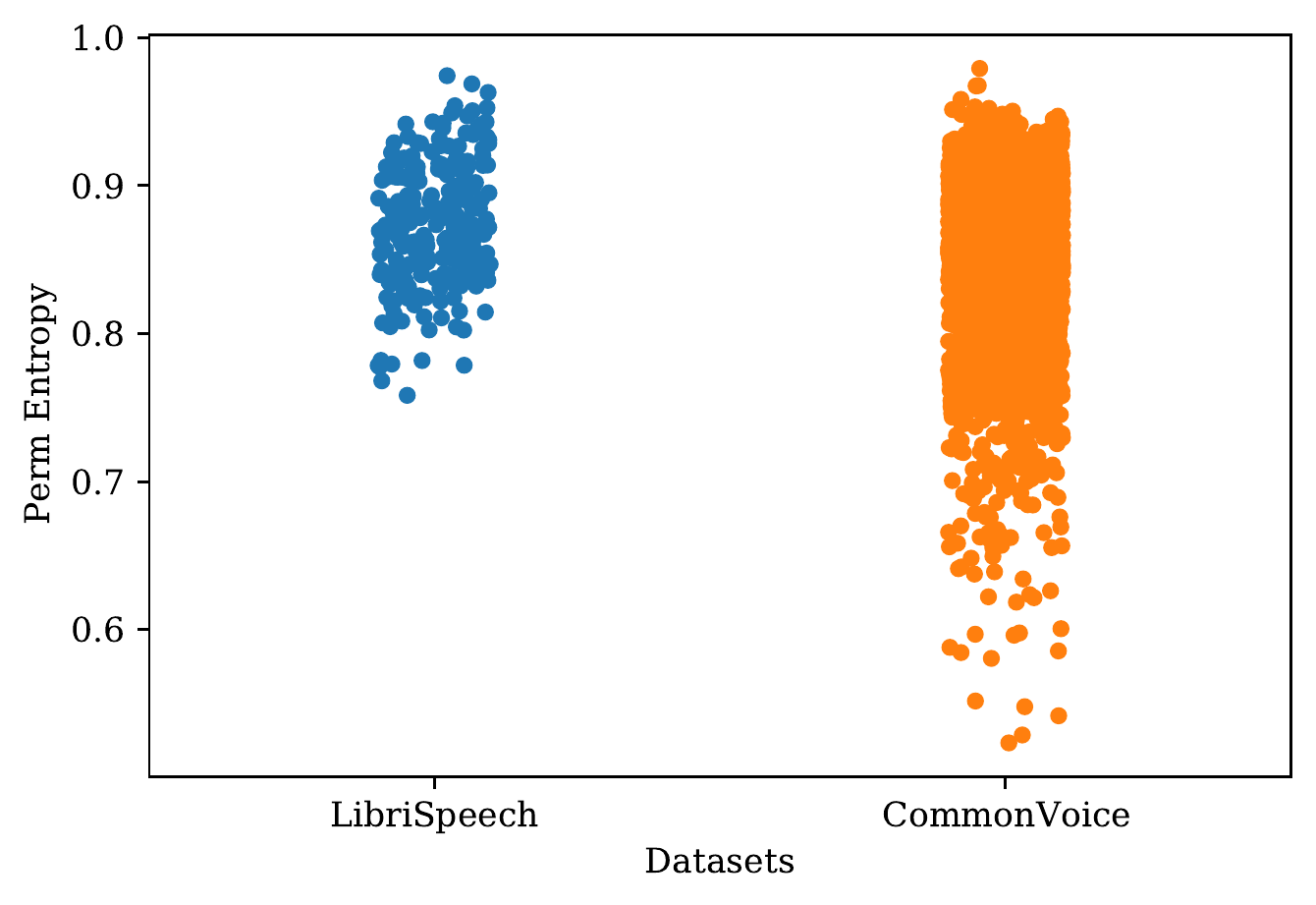}
\end{subfigure}%
\\
\begin{subfigure}{.5\textwidth}
  \centering
  \includegraphics[width=.7\linewidth]{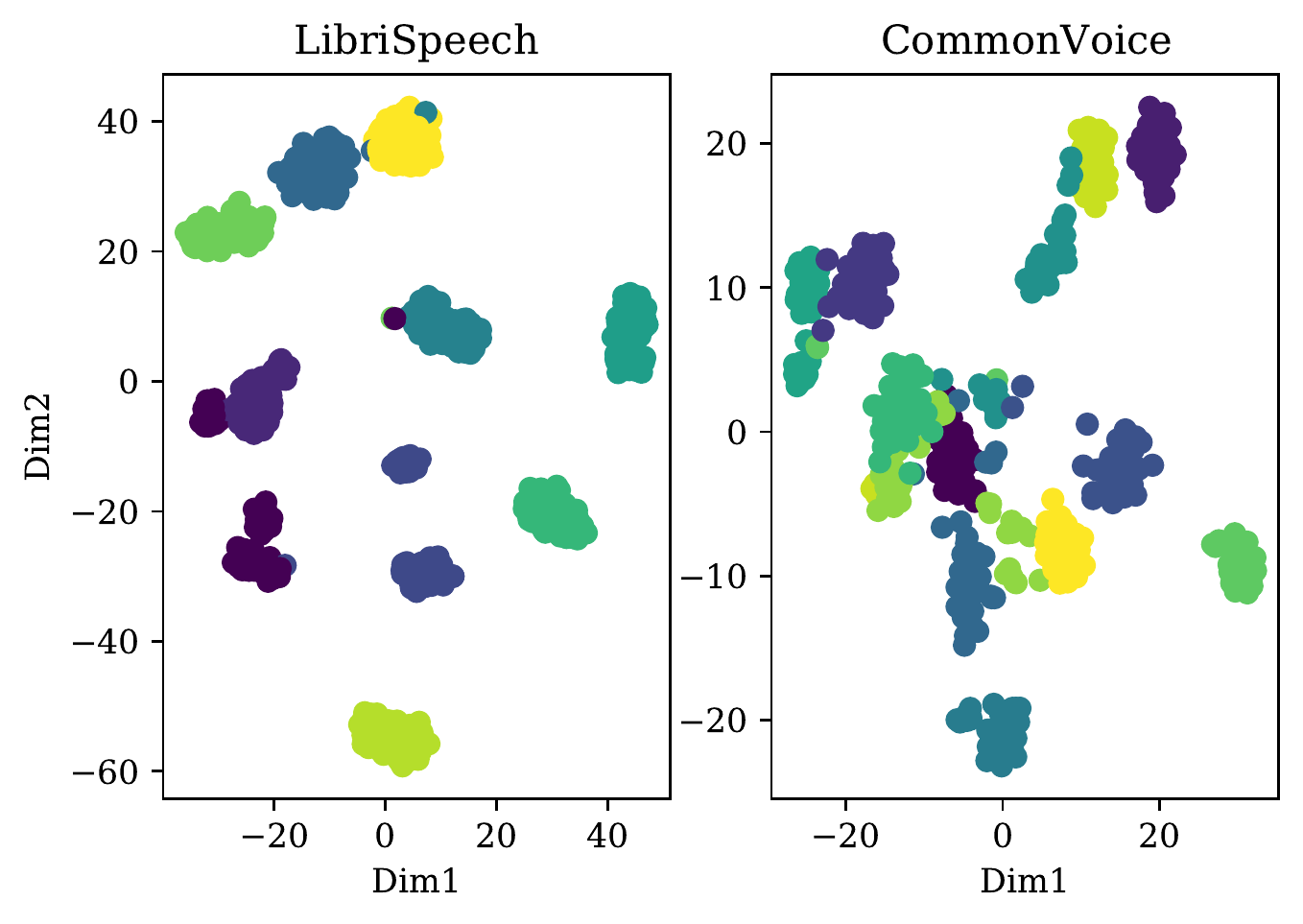}
\end{subfigure}
\caption{\small \textit{(Above)} Strip plot of the permutation entropy mean values per client in Librispeech and Common Voice. CV shows a heavy-tailed distribution as a consequence of the bigger diversity of recording settings. \textit{(Below)} TSNE representation of embedded speech utterances. The colors correspond to the true clients (\textit{i.e.} speakers). CV clients are clearly harder to separate than LS.}
\label{fig:comparison}
\end{figure}

The mean value of the signal feature is computed for every utterance by averaging the per-frame values. Then, for every client we compute the mean value and the standard deviation per client. The former distribution describes the inter-client variation while the latter describes the intra-client one. For the three considered features, the standard deviation of the mean value per client distribution is higher for Common Voice than for LibriSpeech, reaching $0.034$, $11.466$  and  $0.053$ for, respectively, Loudness, logHNR and Permutation Entropy on CV compared to  $0.017$, $9.096$ and $0.040$ on LS. Concerning the intra-client variation, the standard deviation of the standard deviation per client distribution is also higher for CV than for LS reaching $0.009$, $2.69$ and $0.014$ against $0.007$, $2.31$ and $0.007$ respectively for loudness, logHNR and PE. It is also interesting to note the heavy tailed distribution obtained with the Permutation Entropy for CV, as depicted in Fig. \ref{fig:comparison}. Indeed, the Kurtosis reaches $4.16$ on CV versus $-0.13$ for LS. In practice, this mean that many clients may be outliers for CV, drastically impacting FL with conventional aggregation mechanisms. 

\textbf{Blind Signal-to-Noise ratios.}
We further inspect the noise difference between two datasets through computing a blind Signal to Noise Ratio estimation. First, a $10$th-order LPC approximation is computed for every sample. Second, the voiced chunks are detected using the Probabilistic YIN algorithm for F0 estimation \cite{mauch2014pyin}. Finally, considering only the voiced chunks that are simpler to approach with an LPC estimate, the noise in the blind SNR estimation is defined as the difference between the real signal and the LPC approximation. Following the trend observed with the signal features, CV shows a higher standard deviation for the SNR mean values with $\sigma_{SnrCV} = 18.47$ compared to $\sigma_{SnrLS} = 10.32$ for LS. Then, a bigger variation within recordings of the same client is observed. Indeed, the standard deviation of the standard deviations obtained for each audio sample of the same client is higher in CV than in LS, with $6.54$ compared to $3.82$. This suggests a higher variability in the recording conditions with respect to the same client. Common Voice speakers may contribute from different places and devices.

\textbf{Clustering purity.}
We compare the overlap of speakers using pretrained speaker embeddings.
For both datasets, speaker embeddings are computed on each utterance using the \textit{Tristounet} model~\cite{bredin2017tristounet} open-sourced on \textit{pyannote.audio}~\cite{bredin2019pyannoteaudio}. It is important to mention that \textit{Tristounet} is not trained on LS or CV or audio book data. These embeddings are then clustered using the K-means algorithm with \textit{kmeans++} initialization with the number of centroids equal to the number of clients. The purity of the clusters is defined as the proportion of points that belong to the same client as the majority of its computed cluster. Purity reaches $0.77$ on LS and $0.62$ on CV. Fig. \ref{fig:comparison} shows a TSNE representation of the utterance embeddings, and highlights the clustering difficulties in CV. This indicates that CV speakers are harder to separate using speaker embeddings. This confirms the two prior experiments using low-level audio features, as it suggests that varying signal features and recording conditions pollute the speech utterance which leads to harder speaker identification.

The analysis provided in this section evidences the drastic differences in corpora between LS and CV. The latter better captures the complexity that FL systems would face when deployed in the real world.

\section{Experimental Settings}


This section first present the architecture of the E2E speech recognizer. Then, it describes the experimental setup of the FL environment alongside with key hyper-parameters.  

\subsection{E2E Speech Recognizer}
\label{sec:arch}
 
The experiments are based on a Seq2Seq model trained with the joint CTC-attention objective~\cite{kim2017joint}. The encoder follows the CRDNN architecture first described~\cite{SB2021} (\textit{i.e. 2D CNN — LSTM — DNN}). The decoder is a location aware GRU with a single hidden layer. The full set of parameters describing the model are given in the GitHub repository. Models are trained to predict sub-words units. No language model fusion is performed to properly assess the impact of the training procedure on the acoustic models. Data is augmented in the time-domain during training. The model has been implemented within SpeechBrain~\cite{SB2021} and is therefore extremely easy to manipulate, customise and retrain.

\subsection{Realistic Federated Learning}
\label{sec:partition}

Based on the natural partitioning of the CV dataset we conduct two sets of experiments reflecting real usages of FL: 


\begin{figure}[t]
\centering
\scalebox{0.6}{
\begin{tikzpicture}
     \begin{axis}[
             width=0.6\textwidth,
             height=.4\textwidth,
             bar width=.25cm,
             ybar,
             ylabel={Number of clients},
             symbolic x coords={0--10,10--20,20--40,40--60,60--80,80--100,100--150,150--200,200--300,300+},
             xtick=data,
             x tick label style={rotate=45,anchor=east},
             xlabel={Number of samples in each client},
             xlabel style={at={(0.5,-4ex)}}
         ]
         \addplot coordinates {(0--10, 16) (10--20, 96) (20--40, 455) (40--60, 701) (60--80, 293) (80--100, 147) (100--150, 172) (150--200, 82) (200--300, 58) (300+, 43)};
         \addplot coordinates {(0--10, 947) (10--20, 1316) (20--40, 839) (40--60, 386) (60--80, 185) (80--100, 131) (100--150, 127) (150--200, 66) (200--300, 59) (300+, 39)};
         \legend{2K-client, 4K-client}
     \end{axis}
 \end{tikzpicture} 
 }
\caption{\small Illustration of the sample distribution across the 2K-client and 4K-client FL settings from the French Common Voice dataset.}
\label{fig:dis}
\vspace{-0.3cm}
\end{figure}
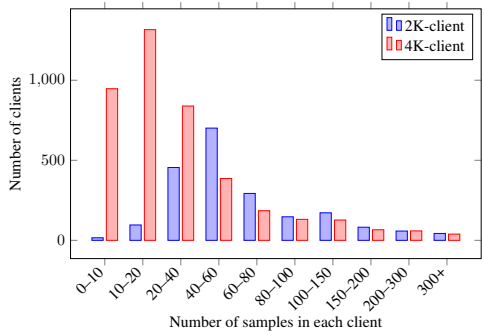

\textbf{\textit{Cross-silo} FL}. In this scenario, clients are generally few, with high availability during all rounds and, often have similar data distribution for training \cite{kairouz2019advances}. 
Shared data is often independent and identically distributed. Our implementation follows that of  \cite{dimitriadis2020federated}, the dataset is split in $10$ random partitions (\textit{i.e.} one per client) with no overlapping speakers each containing roughly the same amount of speech data.

\textbf{\textit{Cross-device} FL}. 
This setup often involves thousands of clients having very different data distributions (non-IID), each participating in just one or a few rounds \cite{kairouz2019advances}. 
Here, we define two settings: First, we simulate a realistic scenario of single speaker using their individual devices. To reproduce this, we naturally divide the training sets based on users into $4095$ and $649$ partitions for French and Italian set, respectively. The second scenario allocates two users per device (e.g as in personal assistants or smart cars). For CV French, this lowers the number of clients to $2036$. As depicted in Fig. \ref{fig:dis}, each setting drastically change the distribution of \textit{low-resources} clients. The $4K$ setup offers a challenging scenario as most clients only contain very few samples. 

\subsection{Federated Learning for ASR: a hybrid approach}
\label{sec:train}

Training E2E ASR models in a FL setting is challenging. Jointly learning the alignment and the latent speech representation is a difficult task that commonly requires large datasets. Therefore, and as we experienced during our analysis, it is nearly impossible to train an E2E ASR model from scratch in a realistic FL setup. Table \ref{tab:result_fr} shows that all the tested existing FL aggregation methods fail to converge without pre-training. This is due to the fact that most of the clients only contain few minutes of speech, resulting in an extremely noisy gradient to learn the alignment from. To overcome this issue we first pre-train the global model on half of the data samples. We do this by partitioning the original dataset into a small subset of speakers (with many samples) for centralised training (referred to subsequently as the \emph{warm-up dataset}) and a much larger subset of speakers (having fewer samples each) for the FL experiment. For CV French, the small subset contains $117$ speakers, leaving the remaining $4095$ speakers for FL. Such statistics are reduced down to $99$ and $649$ speakers for Italian. We argue that this scenario remains realistic as, in practice, centralised data is often available and can therefore be used to bootstrap the alignment. 

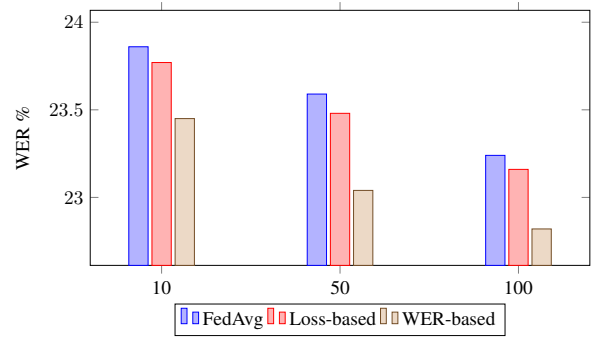
\begin{figure}[!t]
\centering
\scalebox{0.73}{
\begin{tikzpicture}
\begin{axis}[
    width=0.6\textwidth,
    height=.35\textwidth,
    ybar,
    enlargelimits=0.2,
    ylabel={WER \%},
    symbolic x coords={10, 50, 100},
    xtick=data,
    legend style={at={(0.5,-0.15)},
    anchor=north,legend columns=-1},
    every node near coord/.append style={font=\tiny},
    nodes near coords align={vertical},
    ]
\addplot coordinates {(10, 23.86) (50, 23.59) (100, 23.24)};
\addplot coordinates {(10,  23.77) (50, 23.48) (100, 23.16) };
\addplot coordinates {(10, 23.45) (50, 23.04) (100, 22.82) };
\legend{FedAvg,  Loss-based,  WER-based}
\end{axis}
\end{tikzpicture}
}
\caption{\small Speech recognition performance when varying the number of sampled clients per round for the 4K-client setting on French CV.}
\label{fig:k}
\vspace{-0.3cm}
\end{figure}

The number of clients participating in each round influences the outcome of the experiments as well. To quantify this variation, we propose to vary the selected number of clients per round $K$ from $10$ to $100$ for all weighting strategies on the $4K$ set. Then, we simply fix $K$ with respect to the best WER obtained (\textit{i.e.} $100$) for the others setups. For the \textit{cross-silo} environment, all clients are selected every round ($K=10$).

In addition to setting the number of global rounds for the FL experiment, we must define as well the number of local epoch (i.e. on each client). This, however, is a non-trivial task~\cite{mcmahan2017communication}. In practice, we found that increasing the number of local epochs leads to clients over-fitting their own local data.
Hence, clients are locally trained for only $5$ epochs.

Depending on the available compute resources, training concurrently a large number of clients might become challenging. While models may be trained with CPUs or modest GPUs on real embedded hardware (e.g. RapsberryPi or NVIDIA Jetson), our simulated FL setup allows us to run these workloads on modern GPUs (e.g. Nvidia Tesla V100) running multiple clients concurrently on a single GPU and implemented with Flower~\cite{beutel2020flower} and SpeechBrain~\cite{SB2021}.  


Models are finally evaluated both on a centralised test set and at the client-level with a small ensemble of local sentences. Indeed, for the French $4$K setup, each client saves $10$\% (with a minimum of $2$ samples) for testing purposes. To be more specific, centralised speakers are new ones, while local speakers have been seen at training time. 

\begin{table}[t]
\centering
\caption{\small Speech recognition results on the centralised test sets of French (\textit{Fr''}) and Italian (\textit{``It''}) CV dataset for different scenarios and weighting strategies. \textit{``User-based''} FL represents $4$K clients for French and $649$ for Italian.}
\label{tab:result_fr}
\scalebox{0.72}{
\begin{tabular}{l|l|cc}
\toprule
\multicolumn{2}{c}{\textbf{Training Scenario}}           & \multicolumn{1}{l}{ \textbf{Fr WER} (\%)} & \multicolumn{1}{l}{ \textbf{It WER} (\%)} \\ \midrule
             & All data (lower bound) & 20.18 & 17.40                        \\ \cline{2-4} 
Centralised  & 1$^{st}$ half (\emph{warm-up})          & 25.26 & 25.90                        \\ \cline{2-4} 
             & 2$^{nd}$ half (post \emph{warm-up})   & 20.94 & 24.86                       \\ \hline
\multirow{3}{*}{\parbox{2.2cm}{10-clients FL \textit{Cross-silo}}} & FedAvg & 21.26 & 20.97 \\ \cline{2-4} 
             & Loss-based                      & 21.10 & 20.86                        \\ \cline{2-4} 
             & WER-based                        & 20.99 & 19.98                        \\ \hline

\multirow{3}{*}{\parbox{2.2cm}{2K-clients FL \textit{Cross-device}}} & FedAvg  & 22.83 & —  \\ \cline{2-4} 
             & Loss-based                      & 22.67 & —                             \\ \cline{2-4} 
             & WER-based                         & 22.42 & —                             \\ \hline
             
\multirow{3}{*}{\parbox{2.2cm}{User-based FL \textit{Cross-device}}} & FedAvg  & 23.24 & 24.32 \\ \cline{2-4} 
             & Loss-based                       & 23.16  & 24.23                            \\ \cline{2-4} 
             & WER-based                        & 22.82  & 23.86                            \\ \hline
             
\multirow{2}{*}{\parbox{2.2cm}{From scratch}} & FedAvg \cite{mcmahan2017communication}  & 100+ & 100+ \\ \cline{2-4} 
             & FedProx \cite{li2018federated}, FedAdam \cite{reddi2020adaptive}                         & 100+ & 100+                             \\  \bottomrule
\end{tabular}
}
\vspace{-0.3cm}
\end{table}

\section{Speech Recognition Results}
\label{sec:result}

First, we compare the impact of selecting different numbers of clients $K$ per round on the most challenging setup ($4$K clients in French, Fig. \ref{fig:k}). Conversely to the literature , higher values of $K$ tend to produce better WER. This is explained by the heterogeneity of the CV dataset, for which extremely noisy clients may perturb the averaging process with few clients per round. Indeed, few clients remain at more than $100$\% of WER even after the full training. For the remaining of the experiments, $K$ will thus be fixed to $100$.

Table \ref{tab:result_fr} reports the results obtained across the different training setups. We notice that training on the entire dataset in a \emph{centralised} way gives us the best WER with $20.18\%$ and $17.40\%$ for the French and Italian sets respectively, which is comparable to the current best literature \cite{SB2021}. This lower-bound is expected as the system has full visibility of the data and can sample the inputs in an almost IID fashion. On the other hand, when using only the \emph{warm-up} dataset, we notice the effect of having fewer data points for training as the WER increases to $25.26\%$ for the French set and $25.90\%$ for the Italian set. This is expected as well as the system has now less data to learn from. This sheds some light on the inherent lower-bound limitations of FL, limited to partial data observations in each round. The third centralised scenario trains the warmed-up model on the 2nd half of data in an on-line training fashion. This model provides a slightly lower WER compared to all FL models in French set. However, we should note that this is an unrealistic setting as training models in a centralised way would void all the privacy guarantees that FL offers. In particular, this model only gains $0.14\%$ improvement in Italian set compared with the \textit{warm-up} model. This indicates the difficulty of training model on the second half data even in centralised fashion. The results on all FL settings exceed centralised training thanks to the centralised fine-tuning in between each round on the server side.

\begin{figure}[!t]
  \centering
  \includegraphics[width=.8\linewidth]{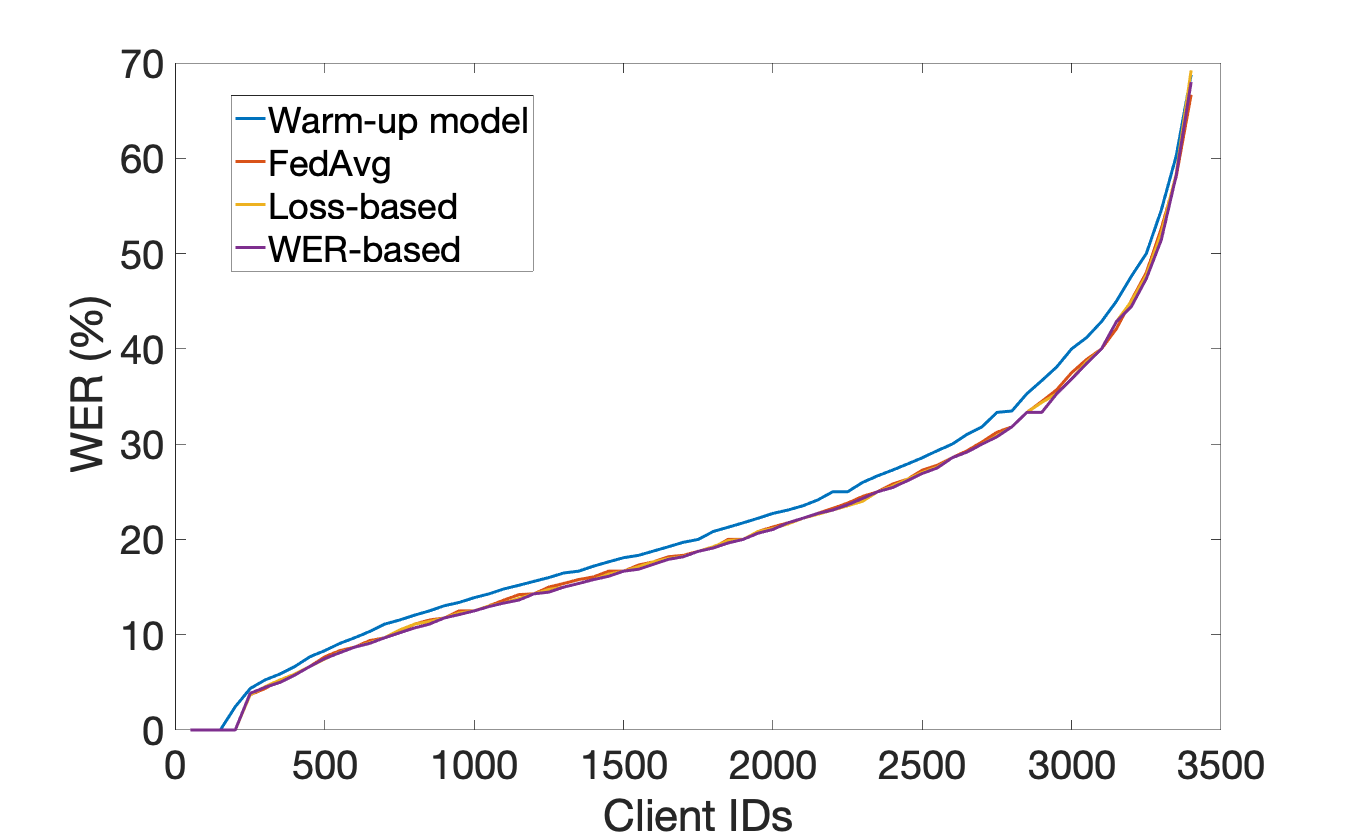}
  \caption{\small Client test performance on the French set of Common Voice for different weighting strategies. The average WER for \textit{warm-up} model, standard FedAvg, loss-based and WER-based aggregation are $23.76$\%, $22.13$\%, $22.11$\% and $21.91$\%. Clients are sorted w.r.t their WER. Clients with a WER higher than $100$\% are removed.}
  \label{fig:cl_wer}
  \vspace{-0.3cm}
\end{figure}

The effect of data visibility can indeed be seen in both \emph{cross-silo} and \emph{cross-device} scenarios, which do not have uniform access to data. However, since this problem is less severe in the former setup, with the correct choice of aggregation strategy we are still able to obtain a WER of $20.99$\% with the French set, which is very close to the centralised lower bound of $20.18$\%. The more challenging Italian set, on the other hand, obtains $19.98\%$ WER with a $2.58\%$ difference to the lower bound. 
As for the \emph{cross-device} scenario, the effect of non-IID data distribution among devices leads to its best WER on French set being $22.43\%$ and $22.82\%$ in the $2$K and $4$K clients settings, even worse ($23.86\%$) with the Italian set. These values are larger than the worst \emph{cross-silo} result, showing the effects of the non-IID nature of the data partitioning.

Compared to different weighting strategies, WER-based and loss-based methods obtain a better performance in all settings, which indicates that weakening the effects of low-quality clients can assist the aggregation process in federated training with heterogeneous data distribution. Herein, we have two types of indicators reflecting the quality of clients. The results in Tab. \ref{tab:result_fr} show that WER-based strategy obtain the lowest WER in both settings.
This could be easily explained by the nature of the strategy which directly optimise the model toward the relevant metric for speech recognition.

Client level test performance is another concern in realistic FL. Fig. \ref{fig:cl_wer} shows the individual WER for each client on French set. All FL methods obtain better performance than the \textit{warm-up} model (blue line), but the difference between the three aggregation strategies becomes less significant. WER-based method, however, obtains the best WER $21.91\%$ when calculating the average performance over all the clients. As previously discussed, we can see that many clients still have a WER higher than $50$\% and $500$ of them even have a local WER higher than $100$\%, clearly indicating the challenging nature of the CV dataset for FL. 

\section{Conclusion}
In this paper, we presented the first study for realistic FL scenarios on attention-based Seq2Seq E2E ASR model with three aggregation weighting strategies -- standard FedAvg, loss-based aggregation and a novel WER-based aggregation. We quantitatively compared LibriSpeech and Common Voice towards a realistic FL setup. All methods were evaluated with \textit{cross-silo} and \textit{cross-device} FL on two languages. 
Our work sets the foundations for future research of realistic FL ASR applications with an open source environment.

\bibliographystyle{IEEEbib}
\bibliography{strings,refs}

\end{document}